\documentclass[apl,twocolumn,floatfix,footinbib,showpacs,superscriptaddress]{revtex4-2} 
\usepackage{graphicx} 
\usepackage{amsfonts,amsmath,amssymb}
\usepackage{amsthm}
\usepackage{dcolumn}
\usepackage{dsfont,bm}
\usepackage[colorlinks=true,linkcolor=blue,pagecolor=blue,filecolor=blue,menucolor=blue,urlcolor=blue,citecolor=blue,anchorcolor=blue]{hyperref}
\usepackage{color}
\usepackage{soul} 
\usepackage{amsbsy}
\usepackage{float}
\usepackage{orcidlink}

\newcommand{\red}[1]{#1}

\begin{document}

\title{Phase jumps in Josephson junctions with time-dependent spin–orbit coupling}
\author{David Monroe \orcidlink{0000-0002-4640-8912}}
\email[\red{Corresponding author: }]{damonroe@buffalo.edu}
\affiliation{University at Buffalo, State University of New York, Buffalo, New York 14260-1500, USA}
\author{Chenghao Shen \orcidlink{0000-0002-8545-6220}}
\affiliation{University at Buffalo, State University of New York, Buffalo, New York 14260-1500, USA}
\author{Dario Tringali \orcidlink{0000-0002-5682-9286}}
\affiliation{University at Buffalo, State University of New York, Buffalo, New York 14260-1500, USA}
\author{Mohammad Alidoust \orcidlink{0000-0002-1554-687X}}
\affiliation{Department of Physics, Norwegian University of Science and Technology, N-7491 Trondheim, Norway}
\author{Tong Zhou \orcidlink{0000-0003-4588-5263}}
\affiliation{University at Buffalo, State University of New York, Buffalo, New York 14260-1500, USA}
\affiliation{Eastern Institute for Advanced Study, Eastern Institute of Technology, Ningbo, Zhejiang 315200, China}
\author{Igor \v{Z}uti\'{c} \orcidlink{0000-0003-2485-226X}}
\email[\red{Corresponding author: }]{zigor@buffalo.edu}
\affiliation{University at Buffalo, State University of New York, Buffalo, New York 14260-1500, USA}
\date{\today}

\begin{abstract}
Planar Josephson junction\red{s} (JJs), based on common superconductors and III–V semiconductors, are 
sought for Majorana states and fault-tolerant quantum computing. However, with gate-tunable spin–orbit coupling (SOC), 
we show that the range of potential applications of such JJs becomes much broader.  
The time-dependent SOC offers unexplored mechanisms for switching JJs, accompanied
by the $2\pi$-phase jumps and the voltage pulses corresponding to the single-flux-quantum transitions, key to high-speed and low-power
superconducting electronics. In a constant applied magnetic field, with Rashba and Dresselhaus SOC, anharmonic current-phase relations, calculated microscopically in these JJs, 
yield a nonreciprocal transport and superconducting diode effect. Together with the time-dependent SOC, this allows us to identify 
a switching mechanism at no applied current bias which
supports fractional-flux-quantum superconducting circuits and neuromorphic computing. 
\end{abstract}
\maketitle

One of the hallmarks of Josephson junctions (JJs) used in their applications is the current-phase relation (CPR)~\cite{Tafuri:2019,Siegel:2012,Soloviev2017:BJN}.
Instead of being driven by the difference of electrical potentials, the dissipationless supercurrent through a JJ, $I$,  is driven by the phase
difference of the superconducting order parameter, $\varphi$, across the junction.    
The dc Josephson effect is commonly given by the CPR, $I(\varphi)=I_c\, \sin\varphi$, where the critical current, $I_c$, is the maximum
supercurrent in a JJ. The resulting ground-state energy corresponds to $\varphi=0$ or integer multiples of $2\pi$.

The control of phase jumps or slips 
in the CPR and different ground states, provides important 
paths in  superconducting electronics and spintronics, neuromorphic computing, advanced qubits, and  fault-tolerant  
quantum computing~\cite{Soloviev2017:BJN,Likharev1991:IEEETAS,Holmes2013:IEEETAS,Yamashita2018:ITE,Giazotto2010:NP,Eschrig2015:RPP,Linder2015:NP,%
Banerjee2014:NC,Crotty2010:PRE,Yamashita2005:PRL,Halterman2015:PRB,Krantz2019:APR,Bal2021:APL,Gungordu2022:JAP,Cai2023:AQT,Amundsen2024:RMP}.
A simple example is the 
$\pi$ junction~\cite{Bulaevskii1977:JETPL} with   
$I(\varphi)=-I_c\, \sin \varphi = I_c\, \sin(\varphi+\pi)$,
with the minimum JJ energy at $\varphi=\pi$~\cite{Yamashita2018:ITE}.  
In superconductor/ferromagnet/superconductor (S/F/S) junctions, 
$0-\pi$ transitions can occur by changing the F-layer thickness~\cite{Valls:2022}.  Replacing F by a spin valve allows implementing 
Josephson magnetic random access memory (JMRAM)~\cite{Birge2019:IEEEML}. 
Unlike MRAM~\cite{Tsymbal:2019}, 
the readout is not based on sensing the tunneling magnetoresistance~\cite{Parkin2004:NM,Ikeda2008:APL}, 
but fast dc-SQUID readout of $\varphi$. Switching the magnetization of a spin valve from parallel to antiparallel orientation
allows tunable $0$ or $\pi$ phase~\cite{Gingrich2016:NP}.

\begin{figure}
\vspace{-2.1cm}
\includegraphics[width=2.3\columnwidth]{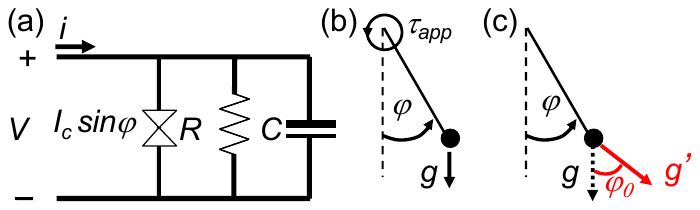}
\vspace{-11.1cm} 
\caption{(a) 
An equivalent Josephson junction (JJ) circuit, driven by bias current, $i$, has normal resistance, $R$, and capacitance, $C$. (b) A pendulum analogy for (a), 
driven by the applied torque, $\tau_\text{app}$, and $\bm{g}$ is the gravitational force. The displacement angle, $\varphi$, is analogous to the superconducting phase 
difference. (c) With spin–orbit coupling and an effective Zeeman field, the effective gravitational force, $\bm{g'}$, is rotated.} 
\label{fig:f1}
\end{figure}

In this work, we study  planar S/two-dimensional electron gas (2DEG)/S JJs, based on
epitaxial Al/InAs heterostructures and motivated by the quest for topological 
quantum computing~\cite{Shabani2016:PRB,Fornieri2019:N,Mayer2020:AEM,Dartiailh2021:PRL,Banerjee2023:PRB}. 
While the experimental detection of $0-\pi$ phase jump~\cite{Dartiailh2021:PRL} confirms topological 
superconductivity in these JJs~\cite{Hell2017:PRL,Pientka2017:PRX,Zhou2022:NC,Zhou2020:PRL}, we explore 
other applications that do not require topological properties, but rely on time-dependent spin–orbit coupling (SOC).  

The feasibility of such time-dependent SOC is demonstrated by
the quasistatic gate-controlled strength in a 2DEG~\cite{Dartiailh2021:PRL,Mayer2020:NC,Gupta2023:NC}. 
Instead of the common form, the CPR can acquire an anomalous phase, $\varphi_0\neq 0, \pi$,
which arises from the broken time-reversal and inversion symmetries~\cite{Reynoso2008:PRL,Buzdin2008:PRL,Konschelle2009:PRL,Sickinger2012:PRL,Strambini2020:NN}.
This example of anomalous Josephson effect in Al/InAs JJs was shown to  be the interplay between 
SOC and the effective Zeeman field $\bm{B}$, with a gate-tunable $\varphi_0$ and generally anharmonic CPR~\cite{Mayer2020:NC,Dartiailh2021:PRL}.
Since the gate switching rates in JJs can exceed the GHz range~\cite{Krantz2019:APR}, the resulting time-dependent electric
field offers an opportunity to implement time-dependent SOC.  

A common  description of a JJ circuit is given in Fig.~\ref{fig:f1}(a) by a Josephson element, resistor, and capacitor, 
using the resistively and capacitively shunted junction model (\red{RCSJ})~\cite{Tafuri:2019}. 
The bias current through the junction, $i$, is the sum of the supercurrent  $\propto \sin\varphi$ and the quasiparticle current flowing in  
the resistor and capacitor. The corresponding phase evolution can be understood from a
pendulum analog in Fig.~\ref{fig:f1}(b).  We generalize it in Fig.~\ref{fig:f1}(c)
to also include the anomalous Josephson effect with $\varphi_0$~\cite{Monroe2022:PRA} and explore
different realizations of $2 \pi$-phase jumps used in high-speed and low-power 
superconducting electronics or neuromorphic computing~\cite{Tafuri:2019,Siegel:2012,Soloviev2017:BJN}.
 
For Al/InAs-based JJs in the ballistic regime their CPR is not simply assumed to be  
$\propto \sin\varphi$ or $\sin(\varphi+\varphi_0)$. Instead, as we discuss below, the CPR is
calculated microscopically from the corresponding Hamiltonian for such junctions 
and we obtain a generalized \red{RCSJ} model
\begin{equation}
d^2\varphi/d\tau^2 + (d\varphi /d\tau)/\sqrt{\beta_c} + I(\varphi, \mu, {\bm h}, \alpha, \beta)/I_c = i/I_c,
\label{eq:pendulum}
\end{equation}
where  
$\tau = \omega_p t$ is a dimensionless time, expressed using the JJ plasma frequency,  $\omega_p = \sqrt{2\pi I_c / \Phi_0 C}$, 
$\Phi_0=h/2e$ is the magnetic flux quantum, and $C$ is the capacitance. The damping is  
given by the Stewart-McCumber parameter, 
$\beta_c = 2\pi I_c C R^2 / \Phi_0$, where  $R$ is the resistance~\cite{Stewart1968:APL,McCumber1968:JAP} and $Q=\sqrt{\beta_c}$ is the quality factor.
The CPR can be modified by the chemical potential $\mu$, and  $\bm{B}$, arising from the applied magnetic field
or magnetic proximity effect~\cite{Zutic2019:MT}. Here, we further show the important dependence of CPR
on Rashba and Dresselhaus SOC in the 2DEG~\cite{Zutic2004:RMP,Pekerten2022:PRB,Pakizer2021:PRR,Scharf2019::PRB}, parametrized
by their strengths $\alpha$ and $\beta$, respectively. 

\begin{figure}[t]
\vspace{-1.2cm}
\includegraphics[width=1.37\columnwidth]{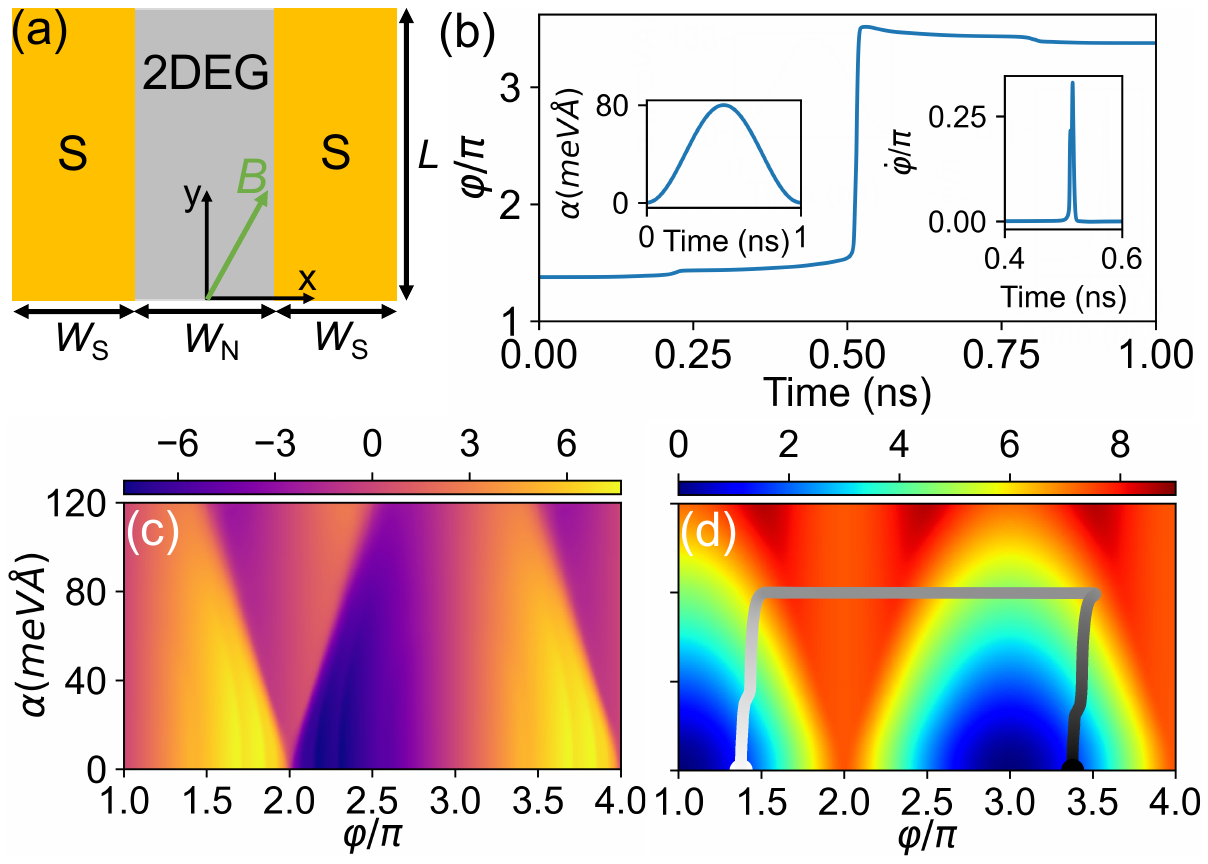}
\vspace{-2.2cm}
\caption{(a) A  JJ schematic. Two superconductors (S) are separated by the 
uncovered 2DEG region with SOC and a Zeeman field $\bm{B}$. (b) Phase evolution, $\varphi(t)$, with sinusoidal
Rashba SOC, $\alpha$ (left) and the resulting voltage pulse (right) at $\omega_p= 1\;\text{THz}$ and $\beta_c=1$. 
The evolution of (c) CPR, normalized by $I_0=2|e\Delta|/\hbar$, and \red{(d)} the JJ energy, normalized by $\Delta$,
as a function of $\varphi$ and  $\alpha$, for chemical potential $\mu = 2.5\Delta$, $B_x=1\;$T, \red{and a critical field $B_c=3\;$T}. 
The white to black path in (d): JJ transition from  $\varphi \approx 1.3\pi$ to $\approx 3.3\pi$ from (b). 
}
\label{fig:f2}
\end{figure}

We calculate CPR for S/2DEG/S geometry with an in-plane field, ${\bm B}$, as 
in Fig.~\ref{fig:f2}(a). The resulting Bogoliubov–de Gennes (BdG) equation~\cite{Valls:2022}, 
$H_\text{BdG}\psi=E \psi$, is expressed using the 
Hamiltonian,  $H_\text{BdG}$, and 
the four-component (in the spin and particle-hole space) wavefunction, $\psi$, for quasiparticle states with energy $E$. 
We solve the discretized BdG equation using a finite-difference method, which
recovers our prior results~\cite{Zhou2022:NC}.
At the boundary, $x=\pm(W_\text{N}/2+W_\text{S}$), $\psi=0$. 
$H_\text{BdG}$  is given 
in terms of the single-particle Hamiltonian, $H_0$,
\begin{equation}
 H_\text{BdG} =  H_0 \tau_z+ (g \mu_B/2)\bm{B}  \cdot \bm{\sigma} + \Delta (x)\tau_+ + \Delta^*(x) \tau_-,
\label{eq:Heff}
\end{equation}
where $H_0$ includes the kinetic and SOC contribution 
\begin{eqnarray}
H_0&=& \frac{\bm{p}^2}{2m^*}-\mu+\frac{\alpha}{\hbar}(p_y\sigma_x-p_x\sigma_y) \\
&+& \frac{\beta}{\hbar}[(p_x\sigma_x-p_y\sigma_y)\cos{2\theta_c} - (p_x\sigma_y+p_y\sigma_x) \sin{2\theta_c}]. \nonumber
\label{eq:H0}
\end{eqnarray}
Here $\sigma_{x,y,z}$ and $\tau_{x,y,z}$ are the Pauli matrices in spin and particle-hole 
space, with $\tau_\pm=(\tau_x \pm i\tau_y)/2$.
$\bm p$ is the in-plane momentum, $m^*$ is the effective electron mass, 
$g$ is the $g$-factor,  and the superconducting pair potential is 
$\Delta(x)=\Delta e^{i \,\text{sgn}(x)\varphi/2} \Theta(|x|-W_\text{N}/2)\Theta(W_\text{S}+W_\text{N}/2-|x|)$, where $\Delta$ is the superconducting gap.
$\theta_c$ characterizes the direction of the current ($x$ axis) with respect to the [100] crystallographic direction of the 2DEG~\cite{Pekerten2022:PRB}.
In our case, we choose $\theta_c=-\pi/4$.
We can then obtain the ground-state JJ energy $E_\mathrm{GS}$ and the CPR,    
$I(\varphi) \propto \partial E_\mathrm{GS}/ \partial \varphi$~\cite{Zagoskin:2014}.

This framework and the intuition from Figs.~\ref{fig:f1}(b) and \ref{fig:f1}(c) allows us to 
\red{examine unexplored} 
methods for JJ switching.
We recall that the ac Josephson effect~\cite{Tafuri:2019}, 
\begin{equation}
V=(\Phi_0/2\pi) \partial \varphi/\partial t,
\label{eq:ac}
\end{equation}
relates  the voltage, $V$, across the JJ in the resistive state with the phase evolution. With a bias current $i\ge I_c$, 
creating torque, $\tau_\text{app} \propto i$, sufficient 
for a full pendulum rotation in Fig.~\ref{fig:f1}(b),  a $2\pi$-phase jump is then accompanied by the voltage pulse corresponding to the area $\int V dt =\Phi_0$ 
and the switching energy $\approx I_c \Phi_0 \sim 10^{-20}\:$J 
at the characteristic frequency, $\omega_p Q/2\pi$. Unlike CMOS logic and older 
operation of JJs~\cite{Siegel:2012}, instead of the voltage, the resulting rapid single-flux-quantum (RSFQ) logic employs the presence (absence) of SFQ as logical ``1" (``0")~\cite{Likharev1991:IEEETAS}. 
\red{In RSFQ logic, a typically large static bias $\sim 0.8I_c$~\cite{Soloviev2017:BJN}, where $\varphi$ is larger than that in Fig.~1(b), implies that
a small time-dependent signal is sufficient for $2\pi$-phase jumps and JJ switching, but requires a large static power dissipation.}

Our Fig.~\ref{fig:f1}(c) reveals a different JJ driving mechanism. The time-dependent Rashba SOC, 
$\alpha(t)$, changed by the gate-controlled electric field, $\bm E$, rotates the direction of the 
gravitational acceleration, $\bm g$ to $\bm g'$.
This could be 
desirable to reduce the static power dissipation in RSFQ logic from the applied current bias~\cite{Soloviev2017:BJN}. 
We assume that a gate control primarily changes $\alpha$, not $\mu$ (nor $\beta$). Experimentally, this could be realized with 
dual \red{gating~\cite{VanTuan2019:PRB,Papadakis1999:S}} to independently tune the carrier density and the electric field.
However, $\alpha(t)$ alone was not sufficient to realize $2\pi$ phase jumps~\cite{Monroe2022:PRA}.
Here, we explore how to overcome this limitation, guided by the 
Al/InAs JJs~\cite{Mayer2020:NC,Dartiailh2021:PRL}.  

Our findings are illustrated for the JJ in Fig.~\ref{fig:f2}(a), 
$W_\text{N}=250\;$nm, $W_\text{S}=300\;$nm, for the normal (N) and S regions, and $L=2\;\mu$m.
 The energies are normalized by $\Delta$ and the supercurrent by
$I_0=2|e\Delta|/\hbar$, where $e$ is the electron charge,
and $|e\Delta|/\hbar$ is the maximum supercurrent in a single-channel short S/N/S JJ~\cite{Zagoskin:2014}.
We use parameters for high-quality epitaxial InAs/Al-based JJs, 
$\Delta_\text{Al}=0.2\;$meV, and $g=10$ for InAs, while its $m^*$ is $0.03$, the electron mass~\cite{Mayer2020:NC,Dartiailh2021:PRL}.
In these JJs, the gate control of Rashba SOC, and thus its magnitude in the range $\alpha \in [0, \, 200 \;\text{meV\AA}$], 
has been  demonstrated~\cite{Mayer2020:NC,Dartiailh2021:PRL}. In III–V 2DEGs the Dresselhaus SOC, $\beta$, can have comparable 
values to $\alpha$ and be gate tunable~\cite{Dettwiler2017:PRX}. Enhanced Al superconductivity is realized near its 2D limit, 
$\Delta_\text{Al}=0.55\;$meV~\cite{vanWeerdenburg2023:SA}. With strong SOC materials next to Al its critical magnetic field can be
much larger, up to $B_c \approx 9\;$T~\cite{Lin2020:PRB,Mazur2022:AM} 
enhancing the JJ parameter space. \red{With these advances, instead of $B_c=1.45\;$T~\cite{Dartiailh2021:PRL} we choose a larger $B_c=3\;$T, but retain 
$\Delta_\text{Al}=0.2\;$meV and its BCS suppression with $B>0$~\cite{Tafuri:2019}}.

In Figs.~\ref{fig:f2}(b)–(d), we study JJ dynamics at   $B_x=1\;$T. From the calculated CPR and JJ energy, we
infer that only a modest energy barrier is preventing 2$\pi$-phase jumps with the evolution of $\alpha(t)$, as indicated with the path in Fig.~\ref{fig:f2}(d),
where the lighter (darker) shades of gray correspond to the earlier (later) times. 
Indeed, we find that for $\bm{E}$-controlled sinusoidal $\alpha(t) \in [0, \ 80 \,\text{meV\AA}$], 
within experimental values, 
a bias 
current $i=0.65 I_c$ is needed to realize reproducible $2\pi$-phase jumps. At larger $\alpha$, $i=0.35  I_c$ is sufficient for $2\pi$-phase jumps. 
As expected from Eq.~(\ref{eq:ac}), these jumps are accompanied by the SFQ voltage 
pulse shown in the inset of Fig.~\ref{fig:f2}(b). 

This scenario, calculated from Eq.~(\ref{eq:ac}), corresponds to the combination of the two JJ driving approaches from Figs.~\ref{fig:f1}(b) and \ref{fig:f1}(c). With $\alpha, B_x, i \neq0$, from the Fig.~\ref{fig:f1}(c) analogy, for this phase evolution we expect $\varphi_0 \neq0$, such that the final $\varphi$ is  
different from $3\pi$. However, 
$B_y=0$ implies that $I(\varphi)=-I(-\varphi)$ and $I(\varphi=0)=0$, as shown in Fig.~\ref{fig:f2}(c). The minima in 
$E_\text{GS}$ in Fig.~\ref{fig:f2}(d) come in pairs at $\pm \varphi_0$~\cite{Alidoust2021:PRB}. The chosen JJ circuit parameters are motivated by 
those measured in the InAs/Al samples,  $I_c \sim 4\;\mathrm{\mu A}$, $R \sim 100\;\Omega$, and $C \sim15\;\mathrm{fF}$, leading to 
$\omega_p \sim900 \;\mathrm{GHz}$ and the damping $\beta_c \sim 1$, which is also suitable for the RSFQ 
applications~\cite{Tafuri:2019,Soloviev2017:BJN,Likharev1991:IEEETAS}. We have verified that the $2\pi$-phase jump is 
insensitive to the specific $\alpha(t)$  details. Nearly identical results and the SFQ pulse are obtained 
for a triangular $\alpha(t)$ of the same amplitude.

From the scenario in Fig.~\ref{fig:f1}(c) and the JJ geometry in Fig.~\ref{fig:f2}(a), we can explore other approaches to switching JJs
which could further reduce the bias current for RSFQ. For example, could the rotation 
of $\bm{B}$ ensure that $\varphi_0\neq0$ (therefore, no minimum $E_\text{GS}$ at $\varphi=-\varphi_0$)
to provide a ratchet effect and a boost for 2$\pi$-phase jumps?

\begin{figure}[t]
\hspace{-0.33cm}
\includegraphics[width=1.022\columnwidth]{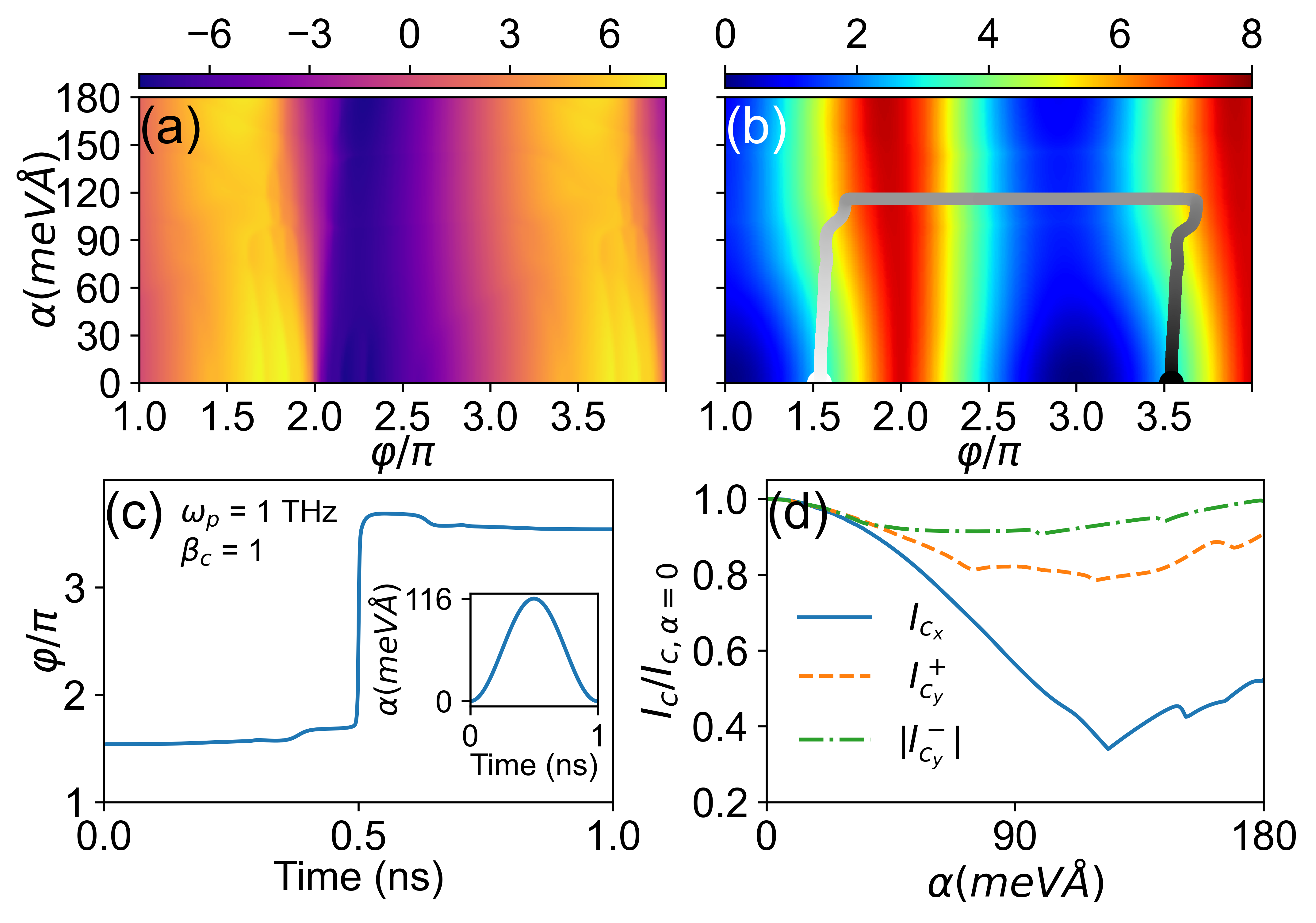}
\vspace{-0.5cm}
\caption{The evolution of (a) CPR, normalized by $I_0$, and (b) the JJ energy, normalized by $\Delta$,
as a function of $\varphi$ and  $\alpha$, for $\mu = 2.5\Delta$ and $-B_y=1\;$T.  
The white to black path in (b) corresponds to the $2\pi$-phase jump, shown also in (c). $\alpha(t)$ changes sinusoidally 
over 1 ns, as in Fig.~\ref{fig:f2}(b), but with the amplitude of $116 \;\text{meV\AA}$. (d) $I_c(\alpha)$ is nonmonotonic 
for a fixed $-B_y=1\;$T. For $B_y$, a Josephson diode effect shows an inequivalence in increasing or 
decreasing $\varphi$, $I_{c_\text{y}}^+ \neq |I_{c_\text{y}}^-|$.
}
\label{fig:f3}
\end{figure}

With Rashba and Dresselhaus SOC, the presence of $\varphi_0$ in the anomalous Josephson
effect can be expressed as~\cite{Alidoust2021:PRB,Alidoust2020:PRB}
\begin{equation}
\varphi_0\propto B_y (\alpha - \beta)(\alpha^2-\beta^2). 
\label{eq:anomal}
\end{equation}
With $\alpha(t)$ and $B_y$, considered in Fig.~\ref{fig:f3}, we can then examine the influence of $\varphi_0$ on the bias current for JJ switching. 
While we again see the 2$\pi$-phase jump in Figs.~\ref{fig:f3}(b) and \ref{fig:f3}(c), the sinusoidal $\alpha(t) \in [0, \; 116] \;\text{meV\AA}$
requires a much higher bias current $i=0.8 I_c$ than  $i=0.65 I_c$ 
from Figs.~\ref{fig:f2}(b) and \ref{fig:f2}(d), at the same magnitude of $\bm{B}$. This is surprising, as the JJ energy landscape in Fig.~\ref{fig:f3}(b)
suggests that the gray path with chosen evolution of $\alpha$ minimizes the local energy barriers.

To understand why such a higher bias current is needed for the anomalous Josephson effect and $\varphi_0$, in Fig.~\ref{fig:f3}(d) we compare the evolution
of $I_c$ with $\alpha$ for a fixed (i) $\bm{B} = B_x \hat{x}$ and (ii) $\bm{B} = B_y \hat{y}$. In the limit $\alpha=0$, as expected, $|\bm{B}| \neq 0$ leads
to the suppression of the critical current, as compared to the zero-field limit. Unlike (i)  with  $I(\varphi)=-I(-\varphi)$ and the  
critical 
current $I_{c_\text{x}}$, independent of the current direction, (ii) is an example of the broken inversion symmetry and the Josephson diode 
effect~\cite{Hu2007:PRL, Amundsen2024:RMP}, where the nonreciprocal transport implies that if $B_y\neq0$ the positive and negative supercurrent has different 
maximum values, $I_{c_\text{y}}^+ \neq |I_{c_\text{y}}^-|$\cite{Yokoyama2014:PRB}. \red{A superconducting diode effect is also possible with a single S region~\cite{Amundsen2024:RMP, Yuan2022:PNAS, Hou2023:PRL, Kochan2023:X}}.

For a ballistic transport of Al/InAs-based JJs with an anharmonic CPR, as shown in Fig.~\ref{fig:f3}(a), the Josephson diode effect was observed 
experimentally~\cite{Amundsen2024:RMP,Dartiailh2021:PRL,Baumgartner2021:NN}. The resulting difference between the critical supercurrent in the forward
and reverse direction can be understood from the process of Andreev reflection at S/N interfaces~\cite{Zutic2004:RMP,Eschrig:2019,Nadgorny:2019} and the 
formation of Andreev bound states~\cite{Kashiwaya2000:RPP,Golubov2004:RMP}  which have unequal contributions right and left moving subgap quasiparticles~\cite{Yokoyama2014:PRB}.
Such ballistic JJs allow for verification of  this prediction for the nonmonotonic  $I_{c_\text{y}}^{\pm}(\alpha)$ in Fig.~\ref{fig:f3}(d).
In contrast, a simple harmonic $I(\varphi)=I_c\, \sin(\varphi+\varphi_0)$, common to an anomalous Josephson effect in a diffusive regime, yields no
diode effect~\cite{Amundsen2024:RMP}.

If we recall Fig.~\ref{fig:f1}(b), a current-driven JJ switching with $2\pi$-phase jumps is achieved for a current exceeding $I_c$ 
(at $B=0$)~\cite{Tafuri:2019}. We then expect that the evolution of the critical current 
in Fig.~\ref{fig:f3}(d) is a guide for the minimum bias current needed
in SOC-driven JJ switching. Indeed, for  (i) and $B_x$, we get $I_{c_\text{x}}(\alpha=80\;\text{meV\AA})=0.65 I_c$, while 
for  (ii) and $B_y$, we get $I_{c_\text{y}}^+(\alpha=116\;\text{meV\AA})=0.8 I_c$, just as we have independently obtained
for the needed bias current by calculating the phase evolution from Eq.~(\ref{eq:pendulum}), which yields $2\pi$ jumps in Figs.~\ref{fig:f2}(b) and \ref{fig:f3}(c). 

Guided by the JJ parameters and the considered effective Zeeman field, our results from Figs.~\ref{fig:f2} and \ref{fig:f3} allow us to establish some trends for the 
SOC-driven JJ switching, relying on the experimentally available range of gate-controlled $\alpha$. If we seek to minimize the bias current for $2\pi$-phase jumps,
which could be both desirable for both low-energy RSFQ and neuromorphic computing~\cite{Soloviev2017:BJN,Crotty2010:PRE},  Fig.~\ref{fig:f3}(d) suggests that
a large SOC and no diode effect are beneficial, while only in a range of $\alpha$ (near 0 and 124$\;\text{meV\AA}$) there is a modest decrease in the 
needed minimum $i$, accompanied by the diode effect. The nonreciprocal transport in the diode effect, $I_{c_\text{y}}^+ \neq |I_{c_\text{y}}^-|$, 
implies that  there is an inequivalence to increase or decrease $\varphi$. In the pendulum analogy from Fig.~\ref{fig:f1}(c), the clockwise and counterclockwise 
$2\pi$-phase rotations are energetically inequivalent, as in the ratchet effect. 

From the above-mentioned discussion, it seems that our goal of a pure SOC-driven JJ switching at $i=0$ is not practical, since it would suggest the need
to tune $\alpha$ to yield a vanishingly small $I_c$. Similarly, seeking to realize the Josephson diode effect only has limited advantages. 
However, there is a need to critically evaluate these assertions as in various III–V semiconductors Dresselhaus SOC strength, $\beta$, can play an important role~\cite{Zutic2004:RMP,Fabian2007:APS}, and as can be seen from Eq.~(\ref{eq:anomal}), the presence of $\varphi_0$ and the implications
of the diode effect could be more complex.

\begin{figure}[t]
\hspace{-0.3cm}
\includegraphics[width=1.02\columnwidth]{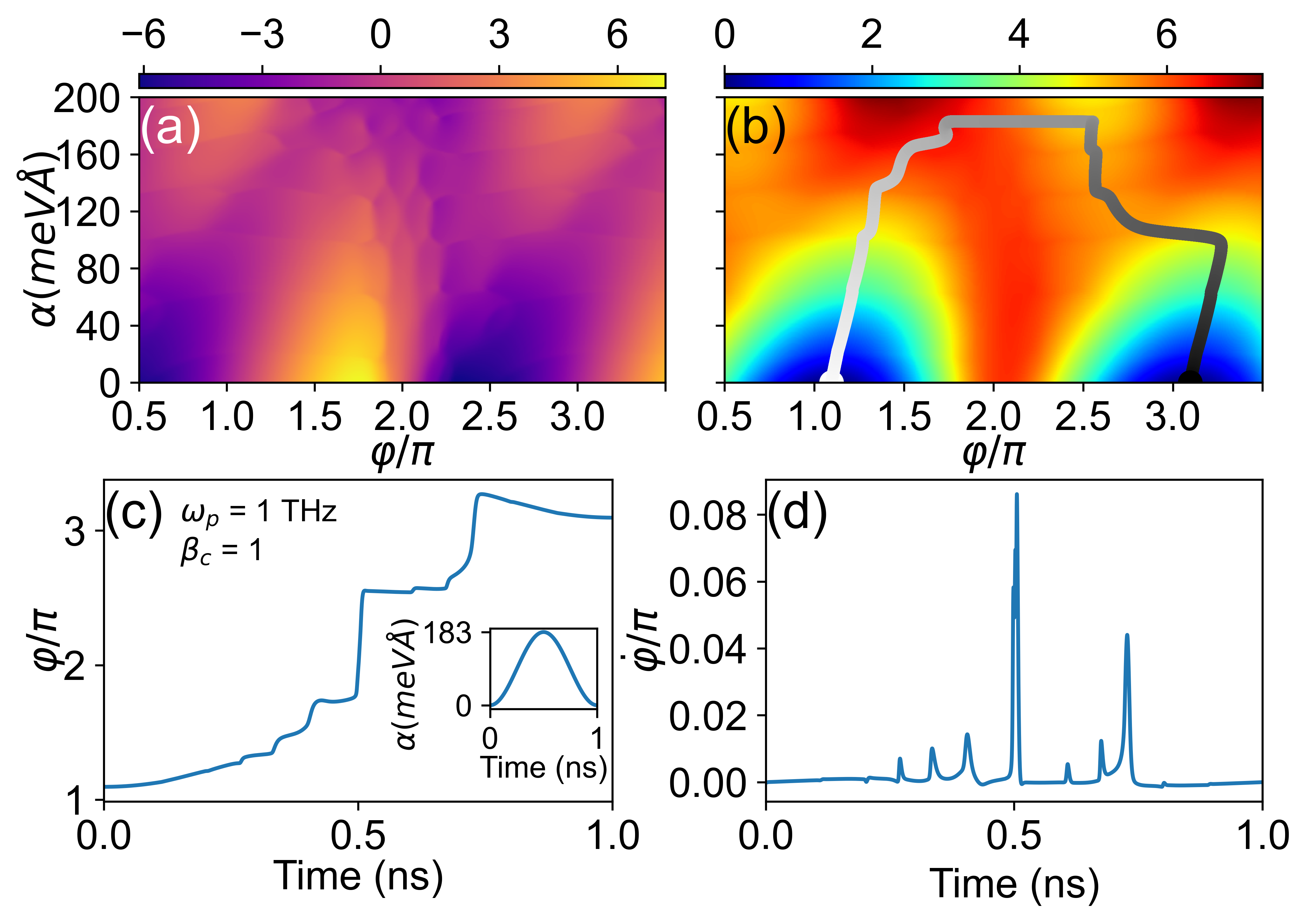}
\vspace{-0.4cm}
\caption{The evolution of (a) CPR and (b) JJ energy with $\varphi$ and $\alpha$, at a fixed $\beta=130\; \text{meV\AA}$, $-B_y=1\;$T, 
and $\mu = 2.5\Delta$. An anharmonic CPR breaks the  $I(-\varphi) = -I(\varphi)$ symmetry in 
(a) and the corresponding anomalous phase, $\varphi_0$,
increases with $\alpha$ in (b). (c) $\varphi(t)$,  for $\omega_p=1\;$THz, $\beta_c=1$,  shows phase jumps near $1.75\pi$ and $2.5\pi$  
with $\alpha(t)$ changing sinusoidally as in Fig.~\ref{fig:f2}(b), but with the amplitude of $183\; \text{meV\AA}$. 
(d) The corresponding voltage pulses, each with a fractional SFQ. 
}
\label{fig:f4}
\end{figure}

To examine this more general situation of the coexisting $\alpha$ and $\beta$, we consider results in Fig.~\ref{fig:f4}, at $B_y\neq 0$,
for a fixed $\beta=130\; \text{meV\AA}$, and $\theta_c=-\pi/4$ in Eq.~(\ref{eq:H0}). 
It is then impossible to combine spatial rotation and spin rotation around $y$ axis to reverse  $I$ and $\varphi$~\cite{Amundsen2024:RMP}.
In Fig.~\ref{fig:f4}(a), this leads to $I(-\varphi)  \neq -I(\varphi)$ and the diode effect, as expected from Eq.~(\ref{eq:anomal}) and $\varphi_0 \neq0$. 

The \red{white to black} 
path in Fig.~\ref{fig:f4}(b) corresponds 
to choosing a harmonic $\alpha(t)$ changing over 1$\;$ns as in Fig.~\ref{fig:f2}(b), but with an amplitude of $183\;\text{meV\AA}$.
Remarkably, the desired $2\pi$-phase jump is now realized at $i=0$, for a pure SOC-driven JJ switching! The energy landscape 
reveals two phase jumps and plateaux, near $\varphi=1.75\pi$ and $2\pi$.  
At $\alpha=0$, the system starts from an energy minimum.  As $\alpha$ increases,  $\beta\neq0$ is responsible for the formation of the 
$\varphi_0 \propto (\alpha-\beta)(\alpha^2-\beta^2)$
state.
The resulting diode effect favors the increase in $\varphi$ (counterclockwise pendulum rotation) and the phase jump near $\varphi=1.75\pi$. 
Subsequently, when $\alpha$ decreases, there is another phase jump near $\varphi=2\pi$ or, equivalently, $\varphi=0$ state, which completes the SOC-driven JJ switching. 

Turning to the phase evolution calculated from Eq.~(\ref{eq:pendulum}) and using the microscopic CPR obtained in Fig.~\ref{fig:f4}(a), we   
see the two main
phase jumps and a plateaux in Fig.~\ref{fig:f4}(c). A manifestation of the resulting ac Josephson effect in Eq.~(\ref{eq:ac}) for such $\varphi(t)$ is the
presence of two voltage pulses shown in Fig.~\ref{fig:f4}(d). Each of them corresponds to the fractional SFQ, while their sum,
along several much smaller peaks from additional small $\varphi$-jumps, yields SFQ. With the push for 
low-energy RSFQ, including using a half-SFQ pulses in more complicated JJ circuits~\cite{Li2021:SST}, it would be important to explore if using the 
fractional-SFQ pulses could be advantageous for the reduced power consumption in superconducting electronics~\cite{Tafuri:2019,Siegel:2012,Soloviev2017:BJN}.

With the current experiments limited to quasistatic SOC changes~\cite{Mayer2020:NC,Dartiailh2021:PRL,Gupta2023:NC}
we have focused on the parameters for the fabricated Al/InAs JJs. However, as the materials 
used in planar JJs expand to also include group IV Ge/SiGe  heterostructures~\cite{Tosato2023:CM} or Au-surface states~\cite{Wei2019:PRL}, 
it would be important to explore other opportunities for SOC-controlled JJ dynamics, since both Rashba and Dresselhaus SOC have not only
different strengths from III–V semiconductors, but also a different form, cubic in the wave vector~\cite{Alidoust2021:PRB,Luethi2023:PRB}.

The chosen GHz change in Rashba SOC is not the ultimate limit for the speed of the SOC-driven JJ switching, but simplifies our calculations~\cite{note}.
In addition to considering a faster gate operation, available in JJ qubits~\cite{Krantz2019:APR}, 
a more rapid phase jumps could be possible. For example, rather that being limited by the changes in the gate-controlled
$\bm{E}$-field and the resulting SOC, the curvature of the JJ energy could lead to an order-of-magnitude faster phase jumps~\cite{Monroe2022:PRA}, while
Fig.~\ref{fig:f4}(d) shows the feasibility of multiple phase jumps and voltage pulses, \red{desirable for enhancing neuromorphic computing~\cite{Tafuri:2019,Crotty2010:PRE}.} 

Beyond our focus on superconducting electronics and neuromorphic computing, in a future work it would be interesting to explore the role 
of time-dependent SOC \red{in} topological superconductivity and fault-tolerant quantum computing~\cite{DasSarma2015:NPJQI,Aasen2016:PRX,Lahtinen2017:SPP}. 
The topological protection offers alternatives to the usual paths of decoupling qubits from their environment~\cite{Krantz2019:APR,Rosen2019:APL}.
Experimental support already reveals an approximately $\pi$-phase jump with an increase in $B_y$ as a signature of 
the transition between trivial and topological superconductivity in SQUID based on two Al/InAs planar JJs~\cite{Dartiailh2021:PRL}. 
With the gate-controlled 
SOC, such a phase jump should also be present at a fixed $B_y$ as an additional contribution to our studied phase dynamics, when extended to  
topological superconductivity. 
Adding multiple gates in this system is predicted to create and reposition Majorana states and therefore probe the non-Abelian
statistics through their fusion~\cite{Zhou2022:NC}, which is crucial for implementing quantum gates and quantum measurements~\cite{Lahtinen2017:SPP,Bonderson2009:AP,Beenakker2020:SP}. Future studies  studies could also explore the role of
time-dependent gate voltage in other implementations of JJs~\cite{Zellekens2022:CP,Fu2024:PRA}.

The data that support the findings of this study are available from the corresponding authors upon reasonable request.

\acknowledgments
We thank Bar{\i}\c{s} Pekerten and Alexander Quinn for valuable discussions. 
This work is supported by NSF ECCS-2130845 (D. M., D. T., and I. \v{Z}), 
US ONR under awards N000141712793 and MURI N000142212764 (C. S. and T.Z.).
Computational resources were provided by the University at Buffalo Center for Computational Research.


\begin{thebibliography}{00}
\bibitem{note}
For GHz switching, the uncertainty in energy is $3\times 10^{-4}\;\text{meV} \ll \Delta_\text{Al}=0.2\;$meV.
\end{thebibliography}
\end{document}